\documentclass[12pt]{iopart}

\usepackage{graphicx}
\usepackage{color}
\usepackage{citesort}
\usepackage{amsthm,amssymb,epsfig,graphics,color,verbatim,booktabs,multirow,rotating}

\begin{document}

%\addbibresource{thebib.bib}

\title[]{Improved magneto-optical trapping of a diatomic molecule}

\author{D.~J. McCarron, E.~B. Norrgard, M.~H. Steinecker \& \mbox{D. DeMille}}

\address{Department of Physics, Yale University, PO Box 208120, New Haven, Connecticut 06520, USA.}
\ead{daniel.mccarron@yale.edu}
\vspace{10pt}
\begin{indented}
\item[]
\end{indented}

\begin{abstract}
We present experimental results from a new scheme for magneto-optically trapping strontium monofluoride (SrF) molecules, which provides increased confinement compared to our original work. The improved trap employs a new approach to magneto-optical trapping presented by M. Tarbutt, \emph{arXiv preprint} 1409.0244, which provided insight for the first time into the source of the restoring force in magneto-optical traps (MOTs) where the cycling transition includes dark Zeeman sublevels (known as type-II MOTs). We measure a radial spring constant $20\times$ greater than in our original work with SrF, comparable to the spring constants reported in atomic type-II MOTs. We achieve a trap lifetime $\tau_{\rm{MOT}}=136(2)$~ms, over $2\times$ longer than originally reported for SrF. Finally, we demonstrate further cooling of the trapped molecules by briefly increasing the trapping lasers' detunings. Our trapping scheme remains a straightforward extension of atomic techniques and marks a step towards the direct production of large, dense, ultracold molecular gases via laser cooling.
\end{abstract}

% Uncomment for PACS numbers
%\pacs{00.00, 20.00, 42.10}
%
% Uncomment for keywords
%\vspace{2pc}
%\noindent{\it Keywords}: XXXXXX, YYYYYYYY, ZZZZZZZZZ
%
% Uncomment for Submitted to journal title message
%\submitto{\JPA}
%
% Uncomment if a separate title page is required
%\maketitle
%
% For two-column output uncomment the next line and choose [10pt] rather than [12pt] in the \documentclass declaration
%\ioptwocol
%

\section{Introduction}
Over the last decade there has been significant and growing interest in the production and manipulation of samples of cold and ultracold polar molecules \cite{Carr2009}. The rich internal structure of molecules coupled with the exquisite control attainable in quantum systems promises new directions for research in diverse fields including ultracold chemistry \cite{Krems2008}, precision measurements \cite{Hunter2012,Tarbutt2013}, and quantum computation \cite{DeMille2002}. The experimental techniques used to produce cold and ultracold molecules span both indirect methods forming molecules from laser cooled atoms, e.g., \cite{Sage2005,Ni2008,Danzl2010,Aikawa2010}, and direct methods ranging from buffer gas cooling \cite{Weinstein1998} and Stark deceleration \cite{Bethlem1999} to evaporative \cite{Stuhl2012} and, notably, laser cooling \cite{Shuman2010}.

Laser cooling and trapping have revolutionized atomic physics, and their maturity and effectiveness for certain atoms may be viewed as the reason indirect methods of molecule production are currently able to attain higher phase-space densities than direct methods. The workhorse technique in cold-atom physics is the magneto-optical trap (MOT), which combines a restoring force from radiation pressure with laser cooling \cite{Raab1987}. Recently our group has achieved a significant milestone in applying laser cooling and trapping techniques to molecules with the realization of the first magneto-optical trap for a diatomic molecule, strontium monofluoride (SrF) \cite{Barry2014}. The power of magneto-optical trapping is apparent, as this work produced the lowest molecular temperature yet by a direct cooling method.

In this work we present an improved SrF MOT. This new MOT uses a trapping scheme proposed by M. Tarbutt \cite{Tarbutt2014}, which differs significantly from that used in our original work \cite{Barry2014}. We demonstrate that, as predicted in ref. \cite{Tarbutt2014}, this new MOT provides increased confinement, allowing access to larger trapped samples of molecules at higher density. Significantly longer trap lifetimes are also observed. Finally, we demonstrate additional in-situ cooling of the trapped molecules by briefly increasing the detuning of the MOT light. The analysis of ref. \cite{Tarbutt2014} provided, for the first time, a qualitative understanding of the mechanisms behind the restoring force in the molecular MOT and other rarely-used atomic MOTs known as type-II (see below). The experimental results described here were motivated by the predictions of ref. \cite{Tarbutt2014}; the agreement of most of our observations with the predictions of ref. \cite{Tarbutt2014} appears to validate the approach of that work. In the next section, we introduce the basic types of MOT, highlight why the source of the restoring force in type-II MOTs was not well understood, and finally outline the new approach to magneto-optical trapping for type-II systems implemented in this work.

\section{Magneto-optical traps and the choice of trapping polarizations}
A magneto-optical trap employs three orthogonal pairs of circularly-polarized laser beams; the lasers are tuned to a frequency slightly below resonance (red-detuned) with an electronic transition in the atomic or molecular species of interest. The Doppler effect and red-detuned light ensure that a particle is more likely to absorb a photon from a beam opposing the particle's velocity, leading to a damping force that cools the particles. To spatially confine the particles in a MOT, it is necessary to apply a 3D magnetic field gradient (quadrupole field) in addition to the laser beams. Here the correct choice of circular polarization for each pair of laser beams is crucial and dependent on the sign of the magnetic field gradient. For the situations discussed in this work, we consider, without loss of generality, the force on particles moving along the z-axis in a magnetic field gradient $dB_{z}/dz$ such that $B_{z}$ is positive when $z>0$. We then consider the effect of different polarizations of the light beam propagating towards the $-z$ direction, for different Zeeman substructures of the states in the trapping transition.

\subsection{Type-I MOTs}
For the common type-I MOT, an $F\rightarrow F'=F+1$ cycling transition is employed, where $F$ is the total angular momentum quantum number and the prime indicates the excited state. This level structure ensures that there are no dark Zeeman sublevels in the ground state. Typically for type-I MOTs, the ground-state magnetic g-factor, $g$, is less than the excited-state g-factor, $g'$. In this case the transition shifted closest to resonance with the red-detuned MOT light is dictated by the sign of the magnetic $g$-factor in the excited state, $g'$. In this common case the correct arrangement of circular polarizations is intuitive and clear. For the nominal case considered here ($z>0$, local magnetic field $B_{z} > 0$, laser beam red-detuned and propagating towards $-z$), when $g'$ is negative a trapping laser should be circularly polarised to drive $\sigma^{+}$ transitions; see Fig. \ref{fig:Fig1}c. If $g'$ is positive the circular polarization should be reversed to drive $\sigma^{-}$ transitions.

\subsection{Type-II MOTs}
The rarely-used type-II MOT operates on an $F\rightarrow F'=F$ or $F\rightarrow F'=F-1$ cycling transition where certain `dark' ground state sub-levels are not coupled to the excited state by the confining laser light. Without a mechanism to `remix' dark states into bright states, such as Larmor precession or optical transitions driven by orthogonal or anti-trapping MOT beams, the scattering rate can tend to zero. Furthermore, the scattering rate from the MOT beam counter-propagating to a particle's displacement from the trap center must be greater than the scattering rate from the co-propagating MOT beam to ensure a confining force. In contrast to the case of type-I MOTs, there has been no widely accepted understanding of the mechanism behind the confining force in a type-II MOT, making the correct choice of circular polarizations in these systems difficult. Despite this lack of understanding, type-II MOTs have been reported in several atomic systems \cite{Prentiss1988,Flemming1997,Nasyrov2001,Tiwari2008}, and these results inspired our group's recent work demonstrating the magneto-optical trapping of a diatomic molecule for the first time \cite{Barry2014}. The rotational degree of freedom in molecules requires the cycling transition of a molecular MOT to correspond to a type-II system \cite{Stuhl2008}.

\subsection{Magneto-optically trapping a molecule - an unexplained restoring force}
In ref. \cite{Barry2014} the confining transitions for the original SrF MOT were selected using the sign of the difference in the magnetic moment between the ground and excited states of the cycling transition. In particular, we chose the circular polarizations such that, for the nominal case, the red-detuned light was as close to resonance as possible for one pair of ground and excited-state Zeeman sublevels in each hyperfine transition. For the $X^{2}\Sigma^{+} \rightarrow A^{2}\Pi_{1/2}$ cycling transition, the ground state $g$-factor is much larger in magnitude than the excited state $g$-factor, i.e. $|g| \gg |g'|$.  Nominally, $g = 2$ for a $^2\Sigma$ state and $g' = 0$ for a $^2\Pi_{1/2}$ state \cite{Brown2003}, though due to spin-orbit mixing effects both values are known to differ from the nominal. We proceeded assuming that $g'\approx0$, resulting in the circular polarizations being dictated by the sign of the $g$-factor for each hyperfine ground state, see Fig. \ref{fig:Fig1}f. Although this was effective at producing a MOT of SrF, similar to atomic type-II MOTs, the source of the small but non-zero restoring force observed was not understood.
\begin{figure}
\centering
\includegraphics[width=10cm]{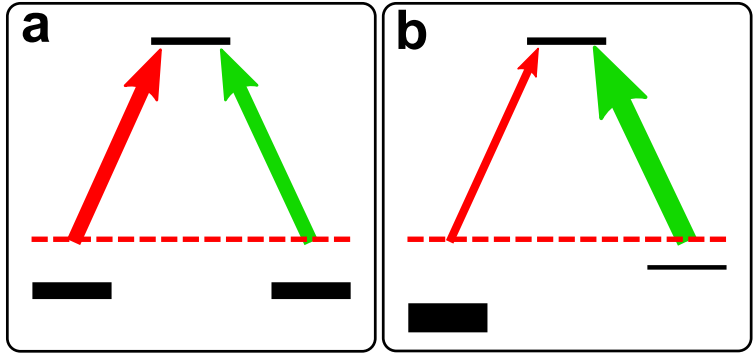}
        \caption{Absence of a restoring force in a Lambda system. Here the red arrow depicts a transition driven by an anti-restoring MOT beam and the green arrow a transition driven by a restoring beam. Arrow widths are proportional to the excitation rate, and each ground-state sublevel line thickness is proportional to its population \textbf{a,} with no magnetic field and \textbf{b,} with a magnetic field applied. The absence of force in both cases (see text) means that the restoring force in type-II MOTs cannot be explained when only the Zeeman splitting in the ground state is considered.}
        \label{fig:FigA}
\end{figure}

To highlight why the source of the restoring force was unclear in type-II systems, consider a lambda system in 1D with a restoring MOT beam driving (for example) $\sigma^{-}$ transitions from one ground-state sublevel and an anti-restoring MOT beam driving $\sigma^{+}$ transitions from the other ground-state sublevel. Assuming that the excited state decays into each ground state with equal probability, and that both beams have the same frequency and intensity, then when both ground states are degenerate (Fig. 1a) the scattering rate is the same for each transition and there is no restoring force. Now consider the case where a magnetic field shifts the sublevel addressed by the restoring beam closer to resonance and the sublevel addressed by the anti-restoring beam further from resonance (Fig. 1b). In this case the excitation rate is smaller from the sublevel further from resonance, so more population accumulates there than in the sublevel closer to resonance. In this picture it can be shown \cite{Minogin1985} that the reduced excitation rate from the sublevel further from resonance, addressed by the anti-restoring beam, is cancelled exactly by the population imbalance between the two sublevels. Hence the scattering rates from the restoring and anti-restoring beams remain equal, and there is no net restoring force for all velocities and any applied magnetic field. We note in passing that the gradient force can be `rectified' and provide confinement in this system for sufficiently high laser intensity and a different polarization configuration \cite{Emile1992}.

\subsection{A proposed explanation}
Numerical simulations recently performed by M.~Tarbutt led him to propose a number of general polarization rules to be followed when magneto-optically trapping \cite{Tarbutt2014}. Crucially, this work proposes that the restoring force present in a MOT relies on the excited state $g$-factor, $g'$, being non-zero. Herein lies a potential explanation of the mechanism behind the confining force in a type-II MOT. The Zeeman splitting in the excited state is essential in order to break the symmetry between confining and anti-confining transitions, allowing a net restoring force to be applied provided that the correct circular polarization is chosen. The larger the magnitude of $g'$, the more this symmetry is broken and the greater the confining force applied. (Surprisingly, this symmetry argument also holds for type-I MOTs where $F<F'$: if $g'=0$ there is no restoring force. However we know of no physical example of such a case since type-I MOTs typically operate on an $nS \rightarrow nP$ transition where $|g|<|g'|$.)

\subsection{When to drive $\sigma^{+}$ or $\sigma^{-}$ transitions}
\begin{figure}
\centering
\includegraphics[width=16cm]{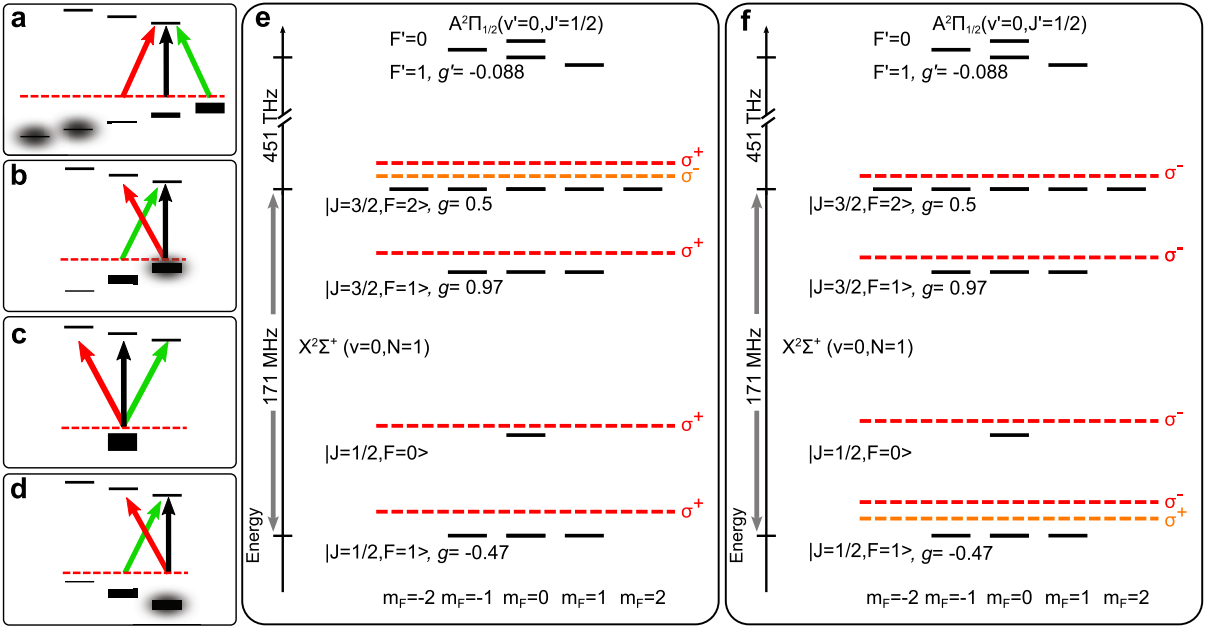}
        \caption{MOT polarizations required to magneto-optically trap on the various hyperfine components of the $X\rightarrow A$ transition in SrF for \textbf{a,} $F=2\rightarrow F'=1$, \textbf{b,} $F=1\rightarrow F'=1$ with positive $g$, \textbf{c,} $F=0\rightarrow F'=1$, and \textbf{d,} $F=1\rightarrow F'=1$ with negative $g$. We do not show transitions to $F'=0$ because they do not contribute towards the confining force. Green arrows mark transitions driven by the confining MOT beam, red arrows mark transitions driven by the anti-confining MOT beam, and black arrows mark transitions driven only by orthogonal MOT beams. Note that orthogonal MOT beams can also drive $\sigma^{+}$ and $\sigma^{-}$ transitions which are not shown. The line thickness of each ground-state sublevel illustrates the population distribution due to decays from the excited state sublevel closest to resonance, $m_{F}'=+1$. Shaded sublevels denote ground states which are dark to the confining MOT beam. \mbox{\rm{\textbf{e,} Improved}} SrF MOT scheme, suggested in ref. \cite{Tarbutt2014} based on the polarization rules outlined in \textbf{a}-\textbf{d}. \textbf{f,} Original trapping scheme \cite{Barry2014} developed assuming that $g'\approx0$. In \textbf{e} and \textbf{f}, dashed lines show the ground state energy with which different laser frequency and polarization components would be resonant.  Red dashed lines show the $\mathcal{L}_{00}$ frequency components and orange dashed lines the $\mathcal{L}^{\dagger}_{00}$ frequency. Ground state Zeeman splittings are not shown as they do not lead to a confining force.}
        \label{fig:Fig1}
\end{figure}

When considering which circular polarization is required to magneto-optically trap, a key consideration is determining which ground-state $m_{F}$ sublevel is populated most often. The $X\rightarrow A$ transition in SrF is an instructive example as it includes all three dipole-allowed changes in angular momenta, namely $F\rightarrow F'=F-1$, $F\rightarrow F'=F$, and $F\rightarrow F'=F+1$ (Figs. \ref{fig:Fig1}a-d). Note that for this transition in SrF, $g'$ is negative; if $g'$ were positive the confining polarizations would be reversed. In addition, since the ground state $g$-factor does not lead to a restoring force, we ignore ground state Zeeman shifts in the ensuing discussion.

The simplest case, $F=0\rightarrow F'=1$ (Fig. \ref{fig:Fig1}c), is that of a type-I MOT with no dark states. Here it is clear that the confining MOT beam should be polarized to drive $\sigma^{+}$ transitions.

For $F=1\rightarrow F'=1$ (Figs. \ref{fig:Fig1}b and \ref{fig:Fig1}d) $\sigma^\mp$ transitions out of $m_{F}=\pm1$ form a pure lambda system and (as discussed above) no restoring force can be applied. However, transitions out of $m_{F}=0$ can lead to a restoring force provided that the restoring MOT beam is polarized to drive $\sigma^{+}$ transitions into the $m_{F}'=+1$ excited state, which is shifted closer to resonance; from here molecules decay into $m_{F}=0,+1$ with equal probability. Although $m_{F}=+1$ is dark to the restoring beam, orthogonal MOT beams can drive transitions out of this sublevel and potentially repopulate $m_{F}=0$, from which the restoring beam can apply a force again. In this scheme the majority of photons scattered are from orthogonal MOT beams; transitions from $m_{F}=0$ are occasionally driven by the restoring beam and rarely by the anti-restoring beam. Additionally, when $m_{F}=-1$ is occasionally populated, transitions can be driven by the restoring MOT beam but not by the anti-restoring beam.

For the final case, $F=2\rightarrow F'=1$ (Fig. \ref{fig:Fig1}a), again the transition closest to resonance is to the $m_{F}'=+1$ sublevel, which decays and populates $m_{F}=+2$ with $60~\%$ probability, $m_{F}=+1$ with $30~\%$ probability, and $m_{F}=0$ with $10~\%$ probability \cite{Metcalf1999}. These branching ratios and the resonance condition highlight that the restoring MOT beam should be polarized to drive $\sigma^{-}$ transitions. The biased branching ratios result in a restoring force in this configuration despite the rarely-visited $m_{F}=-2,-1$ sublevels being dark to the restoring MOT beam and transitions from $m_{F}=0$ being driven more often by the anti-restoring than the restoring MOT beam.

Following the polarization rules developed in ref. \cite{Tarbutt2014} and outlined above, a revised trapping scheme for SrF was proposed in ref. \cite{Tarbutt2014} (Fig. 1e) that differs from that used for the original MOT demonstration in ref. \cite{Barry2014} (Fig. 1f). Simulations of this new scheme in ref. \cite{Tarbutt2014} predicted both tighter confinement and a larger capture volume than those present in the original setup.

\section{Experimental overview}
The SrF MOT presented in this work uses techniques akin to those in atomic MOTs, with a static magnetic quadrupole field and three orthogonal pairs of circularly-polarized laser beams. The optical cycling scheme in SrF on the $X^{2}\Sigma^{+} \rightarrow A^{2}\Pi_{1/2}$ electronic transition has been described elsewhere \cite{Shuman2009,Shuman2010,Barry2012,Barry2014}. In brief, rotational branching is avoided by driving an $N=1(J=3/2,1/2)\rightarrow J'=1/2$ transition \cite{Stuhl2008}, where $N$ is the total angular momentum excluding nuclear and electronic spin and $J$ is the total angular momentum excluding nuclear spin. Highly diagonal Franck-Condon factors suppress vibrational branching, and the calculated vibrational branching fractions $b_{\nu',\nu}$ for the decay of an excited state with vibrational quantum number $\nu'$ to a ground state with vibrational quantum number $\nu$ indicate that only three vibrational repump wavelengths are necessary to scatter $\sim 10^{6}$ photons. Given the measured scattering rate (see below), the three repump lasers, denoted $\mathcal{L}_{10}$, $\mathcal{L}_{21}$, and $\mathcal{L}_{32}$, should give access to trap lifetimes of $\sim1$~s, similar to atomic MOTs. Here the laser addressing the $X(\nu=i)\rightarrow A(\nu'=j)$ transition is labelled $\mathcal{L}_{ij}$ and the detuning of each laser is labelled $\Delta_{ij}$. In addition to the primary trapping laser, $\mathcal{L}_{00}$, there is a secondary trapping laser, labelled $\mathcal{L}_{00}^{\dagger}$ which has polarization orthogonal to the $\mathcal{L}_{00}$ laser; this makes it possible to address each spin-rotation/hyperfine (SR/HF) sublevel with the desired polarization for trapping  (Fig. 1e). Radio-frequency sidebands are added to the $\mathcal{L}_{00}$, $\mathcal{L}_{10}$, $\mathcal{L}_{21}$, and $\mathcal{L}_{32}$ lasers to address the SR/HF structure in the $X^{2}\Sigma^{+}$ state.

In this work, unlike in our previous paper \cite{Barry2014}, we have found that an additional repumping laser, $\mathcal{L}_{00}^{N=3}$, is necessary achieve MOT lifetimes $\tau_{\rm{MOT}}\gtrsim40$~ms. This laser addresses a loss due to excitation to the $A^{2}\Pi_{1/2}(v'=0,J=3/2)$ state, which decays into the dark $X^{2}\Sigma^{+}(v=0, N=3)$ with $30~\%$ probability (the remaining $70~\%$ of decays return to the bright $X^{2}\Sigma^{+}(v=0, N=1)$ state). The $\mathcal{L}_{00}^{N=3}$ laser repumps the $N=3$ population by exciting from $X^{2}\Sigma^{+}(v=0, N=3)$ back to $A^{2}\Pi_{1/2}(v'=0,J=3/2)$, playing a role similar to that of a repump laser in an alkali MOT. More details about the need for the $\mathcal{L}_{00}^{N=3}$ laser are given in section \ref{sec:Lifetime}.

To load the trap, pulses of SrF are produced from a cryogenic buffer gas beam source via laser ablation of an SrF$_{2}$ target at $t=0$~ms \cite{Barry2011} and slowed via radiation pressure from $t=0$~ms to $t=35$~ms \cite{Barry2012,Barry2014}. Trapped molecules are detected using laser-induced fluorescence (LIF) from the $X\rightarrow A$ cycling transition at $\lambda_{00} = 663.3$~nm, imaged onto a CCD. It is not possible to record accurate LIF images of the MOT during the $35$~ms of slowing due to the large amount of scattered light in the trapping region from the slowing beam. All laser powers and beam sizes are the same as those reported in \cite{Barry2014} aside from the addition here of the $\mathcal{L}_{00}^{N=3}$ laser, which has a power of $\approx4$~mW and the same beam size as the other MOT lasers ($1/e^{2}$ intensity diameter of 14~mm).

To modify the original trapping scheme, Fig. 1f, to that shown in Fig. 1e, several simple changes are necessary. First, the type of $\sigma$ transition driven by the $\mathcal{L}_{00}$ and $\mathcal{L}_{00}^{\dagger}$ lasers must be reversed. This is done by either rotating the polarizations of both lasers by $90^{\circ}$ or reversing the orientation of the magnetic field gradient. The frequency of the $\mathcal{L}_{00}^{\dagger}$ laser must also be reduced by $\approx170$~MHz to address the $|J=3/2,F=2\rangle$ manifold. Finally, the modulation frequency of the $\mathcal{L}_{00}$ sidebands is increased from 40.4~MHz to 42.7~MHz so  that the sidebands move to target the lower three SR/HF levels. The frequency of 42.7~MHz minimises the root-mean-squared (r.m.s.) value of the detunings for the lower three SR/HF levels at zero magnetic field ($B=0$~G) and when the $\mathcal{L}_{00}$ laser is tuned nominally to resonance.

Magneto-optically trapped molecules appear as a localized and increased-intensity LIF signal (relative to that of the slowed but untrapped molecular beam) at the center of the trapping region, near the B-field zero. The MOT LIF signal also persists in the trapping region to significantly later times compared to the spatially broad LIF signal from the molecular beam, which almost entirely disappears by $t=65$~ms. The $\mathcal{L}_{00}$, $\mathcal{L}_{00}^{\dagger}$, $\mathcal{L}_{10}$ and $\mathcal{L}_{21}$ lasers are essential to observe the MOT, while the $\mathcal{L}_{32}$ laser is necessary for $\tau_{\rm{MOT}}\gtrsim20$~ms and the $\mathcal{L}_{00}^{N=3}$ laser needs to be present for $\tau_{\rm{MOT}}\gtrsim40$~ms. The magnetic field gradient must satisfy $4$~G/cm$\leq dB/dz\leq 40$~G/cm and the laser slowing must be applied to observe a MOT.

Maximum MOT LIF signals are detected when the laser detunings $\Delta =\Delta_{00}=\Delta_{00}^{\dagger}=- 2\pi\times 10$~MHz $\approx-1.5\Gamma$ ($\Gamma = 2\pi \times 6.6$~MHz is the natural linewidth) and $dB/dz = 9$~G/cm. All repump lasers are tuned to the field-free resonance ($\Delta_{10}=\Delta_{21}=\Delta_{32}=\Delta_{00}^{N=3}=0$) defined as the frequency that gives maximum LIF when light is applied (and retro-reflected) perpendicular to the molecular beam when $dB/dz=0$. All data presented are recorded using these default parameters unless otherwise stated.

\section{Experimental properties of the new MOT}

\subsection{MOT number and density}
\begin{figure}
\centering
\includegraphics[width=12cm]{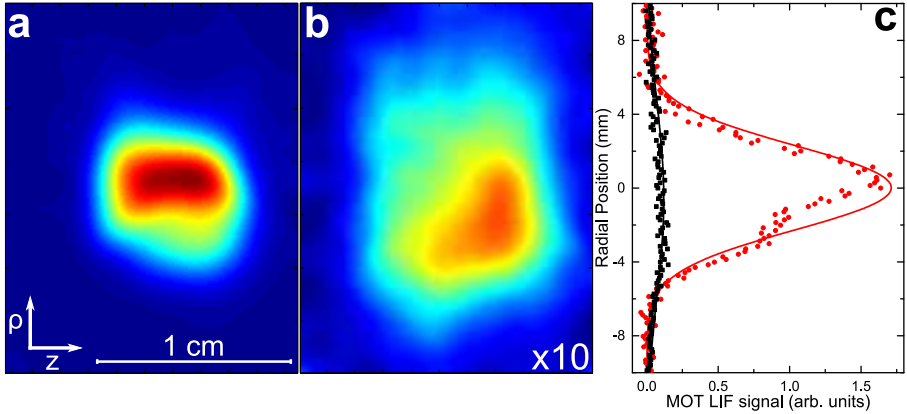}
        \caption{Improved magneto-optical trapping of SrF. MOT LIF images (averaged over 600 pulses) for \textbf{a,} the new trapping scheme and \textbf{b,} the original trapping scheme. The range of the color scale in \textbf{b} has been reduced by $10\times$ compared to the scale in \textbf{a} to ensure that the trapped cloud shape is clearly visible. For both images LIF is measured for a $60$~ms exposure starting from $t=60$~ms, and background counts from scattered light and dark counts are subtracted. \textbf{c,} Radial profiles (through the axial fit centers) of the trapped clouds and 2D Gaussian fits for the new (red) and original (black) schemes highlight the $\sim5\times$ increase in LIF. LIF images are smoothed with a Gaussian of width $\sigma=0.7$~mm, but all fits and analysis are performed on unsmoothed data.}
        \label{fig:LIFcomp}
\end{figure}

Several features of the MOT produced with the new scheme are strikingly different from those using the original scheme. For example, MOT LIF images show that the trapped cloud is spatially smaller than for the original trapping scheme \cite{Barry2014} (Fig. \ref{fig:LIFcomp}). A two-dimensional Gaussian fit to the LIF intensity profile gives radial and axial r.m.s. widths of $\rho_{\rm{rms}}=1.9(1)$~mm and $z_{\rm{rms}}=2.1(1)$~mm respectively, which may be compared to values $\rho_{\rm{rms}}=4.1(1)$~mm and $z_{\rm{rms}}=2.6(1)$~mm previously. This indicates that the confining and/or damping forces within the trap have changed.

The spontaneous scattering rate, $R_{\rm{sc}}$, is measured using the same method presented in ref. \cite{Barry2014}. In short, the $\mathcal{L}_{21}$ laser is extinguished at $t=58.6$~ms and the LIF decay constant, $\tau_{\nu=2}$, is measured as molecules are optically pumped into the now dark $X^{2}\Sigma^{+}(\nu=2)$ state. We measure $\tau_{\nu=2}=0.74^{+0.34}_{-0.27}$~ms and, using the vibrational branching fraction $b_{02}\approx0.004$, find that $R_{\rm{sc}}\approx3.8^{+2.2}_{-1.2}\times10^{6}$~s$^{-1}$. This is not significantly different from the scattering rate in the original configuration. The detection system is the same as used in ref. \cite{Barry2014}, with a measured efficiency of $\sim\!0.8~\%$; the measured value of $1.1\times10^{6}$ detected photoelectron counts imaged at $t=100$~ms, with camera exposure $\Delta t_{\rm{exp}}=60$~ms, corresponds to a trapped population at the end of the slowing period (i.e. at $t=35$~ms) of $N_{\rm{MOT}}\approx500$ molecules, roughly $1.7\times$ greater than in ref. \cite{Barry2014}. The peak density is $n_{\rm{MOT}}\approx4000$~cm$^{-3}$, roughly $7\times$ greater than that measured in ref. \cite{Barry2014}.

\subsection{Spring constant and damping force}
\begin{figure}
\centering
\includegraphics[width=10cm]{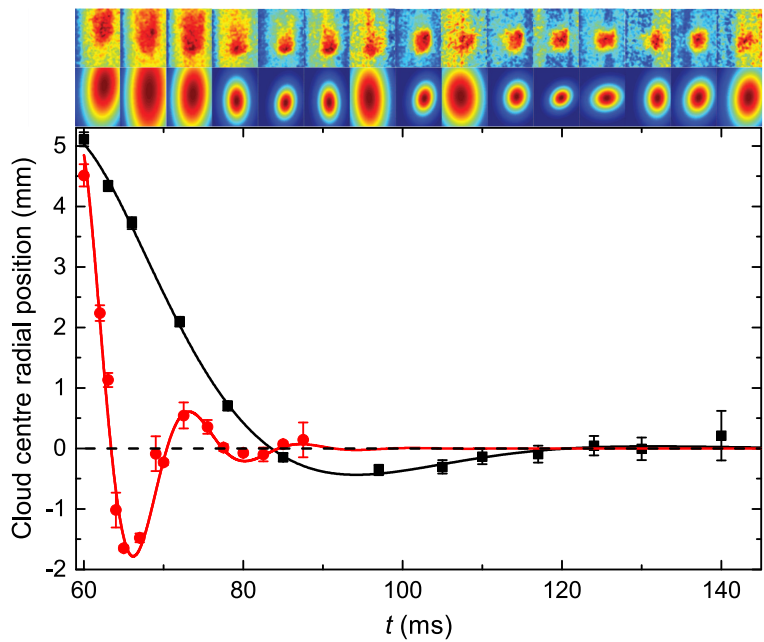}
        \caption{The trapped cloud's response to a rapid displacement of the trap center for the new (red circles) and original (black squares) trapping schemes. Top, MOT LIF images using the new scheme averaged over 2200 pulses (images are ordered chronologically and span $t=60\rightarrow 88$~ms); middle, 2D Gaussian fits to LIF images; bottom (main panel), extracted radial MOT center as a function of time. The fits are to the motion of a damped harmonic oscillator, and zero is set to the position of the unbiased MOT center. For clarity, LIF images are smoothed with a Gaussian of width $\sigma=0.7$~mm. Error bars show the $\pm1\sigma$ confidence interval from a $\chi^{2}$ analysis of the fit, and all fits and analysis are performed using the unsmoothed data.}
        \label{fig:Oscillation}
\end{figure}

To directly probe the confining and damping forces within the trap, the MOT response to a rapid displacement of the trap center is measured. During the loading phase, a magnetic bias field offsets the trap center $\approx5$~mm radially, along the axis of the molecular beam. The bias field is then rapidly switched off to release the trapped molecules into the unbiased potential. To avoid significant LIF from the slowed molecular beam, the displaced MOT is released at $t=59$~ms. The molecular cloud's position is measured as a function of time using short camera exposures, $\Delta t_{\rm{exp}}=2$~ms, and compared to data taken using the original trapping scheme \cite{Barry2014} as shown in Fig. \ref{fig:Oscillation}.

The cloud moves with damped harmonic motion about the trap center with radial frequency $\omega_{\rho}=2\pi \times 76(5)$~Hz and damping coefficient $\alpha/m_{\rm{SrF}} = 310(30)$~s$^{-1}$, where $m_{\rm{SrF}}$ is the mass. The new trapping scheme used in this work produces a potential with trap frequencies over $4\times$ greater than those measured in our original trap, where $\omega_{\rho}=2\pi \times 17.2(6)$~Hz \cite{Barry2014}, indicating that the trap spring constants ($\kappa_{\rho,z}=m_{\rm{SrF}}\omega_{\rho,z}^{2}$) are 20$\times$ greater. In addition to tighter confinement, the measured damping coefficient is increased by a factor of 2.

\subsection{MOT temperature}
Once the trap frequency and cloud size are known, the equipartition theorem can be used to calculate the radial MOT temperature, $T_{\rho}$. We find $T_{\rho} = m_{\rm{SrF}}\omega_{\rho}^{2}\rho_{\rm{rms}}^{2}/k_{B} = 11(2)$~mK, where $k_{B}$ is the Boltzmann constant. Assuming $\omega_{z} = \sqrt{2} \omega_{\rho}$, which applies to atomic MOTs in quadrupole magnetic fields \cite{Metcalf1999}, the calculated axial trap frequency $\omega_{z}=2\pi \times 107(7)$~Hz and measured MOT axial width, $z_{\rm{rms}}$, correspond to an axial temperature $T_{z}=26(3)$~mK. The MOT temperature extracted using this method is $T_{\rm{MOT}} = (T_{\rho}^{2}T_{z})^{1/3}= 14(5)$~mK.

An independent measurement of the MOT temperature is made using the release-and-recapture method. In this method trapped molecules are released from the trap for a variable time of flight $\Delta t_{\rm{TOF}}$ before the trap is switched back on; this recaptures a fraction of the initial molecules that depends on their mean velocity and therefore temperature \cite{Lett1988}. To avoid LIF from the slowed molecular beam, the MOT is released (by extinguishing the $\mathcal{L}_{21}$ laser) at a fixed time of $t=t_{\rm{rel}}=100$~ms and the time of flight $\Delta t_{\rm{TOF}}$ is varied from $0$ to $30$~ms. After each release, the $\mathcal{L}_{21}$ laser is turned on at $t=t_{\rm{rel}}+\Delta t_{\rm{TOF}}$ to recapture the remaining molecules, and imaging begins at $t=t_{\rm{rel}}+\Delta t_{\rm{TOF}}+3$~ms using $\Delta t_{\rm{exp}}=60$~ms (Fig. \ref{fig:Cooling}).

To extract the MOT temperature, the measured recaptured fraction as a function of $\Delta t_{\rm{TOF}}$ is compared to a Monte Carlo simulation. The model assumes a spherical trap volume with radius $r_{\rm{cap}}$; molecule velocities are drawn from an isotropic Boltzmann distribution and the effects of gravity are included. Molecules inside the trap volume are assumed to be recaptured with $100~\%$ efficiency while those outside are assumed to be lost. MOT LIF images are used to infer the initial spatial distribution of the MOT prior to release, assuming that the MOT is radially symmetric. While the uncertainty in $r_{\rm{cap}}$ limits the absolute accuracy of extracted temperatures, it does not influence comparative measurements between data sets to detect changes in temperature, provided that the MOT parameters during recapture are fixed. We fix $r_{\rm{cap}}=d_{\lambda}/2$, where $d_{\lambda}=23$~mm is the MOT beam diameter, to obtain an upper limit on the isotropic temperature, $T_{\rm{iso}}$. We find $T_{\rm{iso}}<12(2)$~mK, in good agreement with the MOT temperature derived from the MOT oscillation measurement.

The MOT temperature is increased by $\sim\!5\times$ compared to the SrF MOT formed with the original trapping scheme and described in ref. \cite{Barry2014}, where \mbox{$T_{\rm{MOT}}=2.3(4)$~mK}. The damping coefficient measured for the new configuration in this work indicates an increased cooling rate, but evidently this increased cooling is overwhelmed by an increase in the heating rate. The increased heating could conceivably be due to the decreased MOT size and hence increased mean laser intensity experienced by a trapped molecule. This change is not reflected in the measured value of $R_{\rm{sc}}$, so that the increased heating cannot be explained by an increase in spontaneous emission random-walk heating, but there may be additional heating mechanisms that are sensitive to laser intensity or to the particular mix of polarizations and frequencies applied \cite{Gordon1980}. Unfortunately, no estimate is given in ref. \cite{Tarbutt2014} of the temperature or damping force for the new configuration.  However, we note that, in ref. \cite{Tarbutt2014}, the calculated damping coefficient for the original MOT configuration exceeded the measured value in ref. \cite{Barry2014} by over an order of magnitude.  Thus, it appears that both the heating and cooling rates in these type-II MOTs remain poorly understood.

\subsection{MOT lifetime} \label{sec:Lifetime}
\begin{figure}
\centering
\includegraphics[width=10cm]{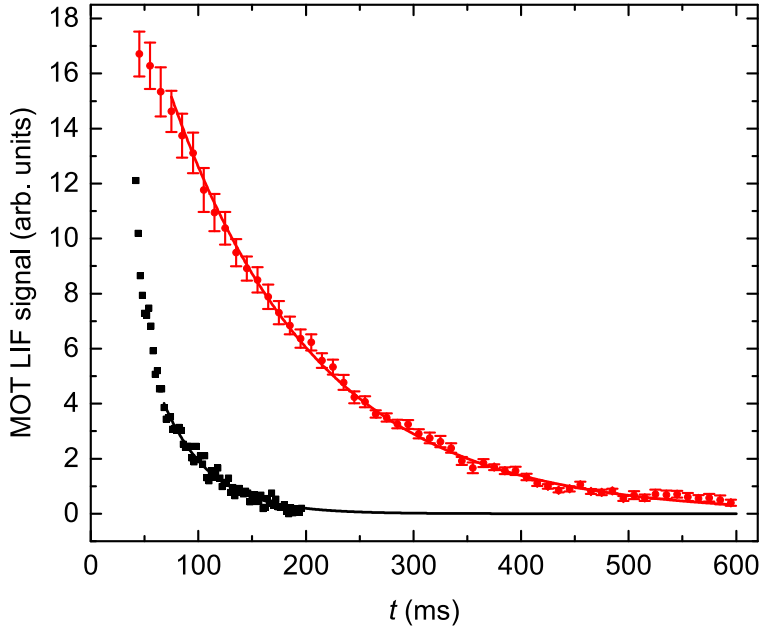}
        \caption{MOT LIF in the trapping region as a function of $t$ using the new (red circles) and original (black squares) trapping schemes. The lines show single exponential decays fit from $t=70$~ms to avoid LIF from the slowed but untrapped molecular beam and give $\tau_{\rm{MOT}}=136(2)$ and $\tau_{\rm{MOT}}=56(4)$ for the new and original trapping schemes respectively. The $\mathcal{L}_{00}^{N=3}$ laser is present for the data taken using the new trapping scheme. Error bars show the $\pm1\sigma$ confidence interval.}
        \label{fig:Lifetime}
\end{figure}

The MOT lifetime, $\tau_{\rm{MOT}}$, is determined by measuring MOT LIF in the trapping region as a function of time. Here we measure $\tau_{\rm{MOT}}=136(2)$~ms. This corresponds to 12 trap periods and is over $2\times$ longer than the trap lifetime measured for the original trapping scheme, where $\tau_{\rm{MOT}}=56(4)$~ms (Fig. \ref{fig:Lifetime}). With the original trapping scheme, $\tau_{\rm{MOT}}$ was observed to be strongly dependent on the MOT beam diameter ($d_{\lambda}$), and we concluded that the dominant loss mechanism was `boil-off' due to the low ratio of trap depth to MOT temperature \cite{Barry2014}. With the new trapping scheme, we measure $\tau_{\rm{MOT}}$ to be largely independent of $d_{\lambda}$, indicating that the trap depth to MOT temperature ratio has increased and is no longer a limiting factor in the trap lifetime.

An upper limit on the trap depth, $U^{\rm{max}}_{\rm{MOT}}=\frac{1}{2}\kappa_{\rho}(d_{\lambda}/2)^{2}$, is estimated assuming that $\kappa_{\rho}$ is constant to the edges of the MOT beam. This gives $U^{\rm{max}}_{\rm{MOT}}/k_{B}= 190(30)$~mK $\approx 16~T_{\rm{MOT}}$. The ratio of trap depth to MOT temperature is thus increased by $4\times$ compared to \cite{Barry2014} but remains in stark contrast to atomic MOTs where $U_{\rm{MOT}}/k_{B}\approx 10^{3}~T_{\rm{MOT}}$.

Additional measurements were made to determine whether the observed lifetime was limited primarily by scattering from background gas in our chamber. To do this, the MOT lifetime was measured as a function of background pressure by varying the the pumping speed in the trapping region while the He flow rate from the beam source was fixed at 5 standard cubic centimeters per minute (sccm). By extrapolating a linear fit to the total loss rate vs. pressure data down to our typical operating pressure of $\approx2\times10^{-9}$~torr, we estimate that the loss rate due to collisions with background He, $\Gamma_{\rm{He}}=0.9(4)$~s$^{-1}$, is only a small fraction of the current total loss rate $1/\tau_{\rm{MOT}}=7.4(1)$~s$^{-1}$. Hence reducing the background pressure will not significantly improve the MOT lifetime. We note in passing that the MOT lifetime is also not sensitive to ballistic He from the cryogenic buffer gas source: it changes negligibly for increased flow rates up to $20$~sccm.

In the absence of the $\mathcal{L}_{00}^{N=3}$ rotational repumping laser, the MOT lifetime was observed to be $\tau_{\rm{MOT}}=39(1)$~ms, shorter than the lifetime measured using the original trapping scheme \cite{Barry2014}. Given the increased trap depth to MOT temperature ratio, and measured independence of $\tau_{\rm{MOT}}$ on $d_{\lambda}$, this result was surprising. While searching for the dominant loss mechanism from the new trap, a $60~\%$ decrease in the $\mathcal{L}_{00}$ laser intensity was found to increase the MOT lifetime by $\approx50~\%$, indicating that the dominant loss mechanism was related to scattering photons. Adding the $\mathcal{L}_{00}^{N=3}$ laser (while at full $\mathcal{L}_{00}$ intensity) to repump population from the dark $X^{2}\Sigma^{+}(v=0, N=3)$ state increased the MOT lifetime by $\sim4\times$, to its current value.

To understand the rate of pumping into the $N=3$ state, we considered the effect of off-resonant pumping into this state by the trapping lasers. The $A^{2}\Pi_{1/2}(v'=0,J=3/2)$ excited state, which decays into the \mbox{$X^{2}\Sigma^{+}(v=0, N=3)$} ground state with \mbox{$30~\%$} probability, lies 17~GHz above the excited state of our cycling transition \cite{Sheridan2009}. The measured value of $R_{\rm{sc}}$ can be used to calculate the rate of off-resonant excitation to the $A^{2}\Pi_{1/2}(v'=0,J=3/2)$ state, $R_{\rm{sc}}(\Delta)$. We find \mbox{$R_{\rm{sc}}(\Delta=17~\rm{GHz})= 1$}~s$^{-1}$, corresponding to $\tau_{\rm{MOT}}\approx3$~s. Clearly this effect cannot explain the observed loss rate.

A plausible explanation for the pumping mechanism is that our trapping laser is contaminated by broadband amplified spontaneous emission (ASE), with non-negligible spectral density at the frequency of the $X^{2}\Sigma^{+}(v=0, N=3) \rightarrow A^{2}\Pi_{1/2}(v'=0,J=3/2)$ transition. The type of semiconductor tapered amplifier used to boost the $\mathcal{L}_{00}$ laser power is known to be susceptible to emitting ASE. Although preliminary measurements using the original trapping scheme showed no evidence of significant loss into $N=3$, this loss mechanism may have contributed towards limiting the final MOT lifetime reported in our previous work \cite{Barry2014}. However, limits on the loss into $N=3$ from preliminary data indicate that the loss rate was less then than it is now. This is not inconsistent with the proposed explanation, since the level of ASE from a tapered amplifier can change due to aging of the gain medium and/or minor changes in alignment and hence could be larger now than before. While this loss mechanism is not intrinsic to our physical system, it may occur in other experiments that use similar laser technology, so we highlight the issue with the hope of helping others avoid the same problem in the future.

\subsection{Comparison of MOT properties}
 Table \ref{table:Parameters} compares the MOT parameters we have determined in this work using the new configuration suggested by Tarbutt \cite{Tarbutt2014} and in our original observation of a molecular MOT \cite{Barry2014}.
\begin{table} \label{table:Parameters}
\begin{tabular}{| l | c | c | l |}
\hline
Parameter	 & This work & Original work	& Unit \\
\hline
Number of molecules, $N_{\rm{MOT}}$	 & 500 &	300	& \\
Density, $n_{\rm{MOT}}$ &	4000	 & 600	& $\rm{cm}^{-3}$ \\
Temperature, $T_{\rm{MOT}}$	& 12(2) & 2.5(5) 	& mK \\
Radial trap frequency, $\omega_{\rho}$ &	$2\pi\times76(5)$ &	$2\pi\times17.2(6)$ &	Hz\\
Damping coefficient, $\alpha/m_{\rm{SrF}}$ &	310(30)	& 140(10)	& $\rm{s}^{-1}$ \\
Optimal detuning, $\Delta$ &	$-2\pi\times10$ &	$-2\pi\times8$	& MHz \\
Optimal gradient, $dB/dz$	& 9	& 15	& $\rm{G}/\rm{cm}$ \\
Gradients with observed MOT &	4--40	& 4--30	& $\rm{G}/\rm{cm}$ \\
$U^{\rm{max}}_{\rm{MOT}}/k_{B}T$	& 16	& 4	& \\
Lifetime, $\tau_{\rm{MOT}}$	& 136(2) &	$56(4)^{*}$ &	ms \\
Scattering rate, $R_{\rm{sc}}$	& $3.8^{+2.2}_{-1.2}$ &	$4.3^{+4.1}_{-2.2}$ &	$\rm{s}^{-1}$ \\
Radial width, $\rho_{\rm{rms}}$ &	1.9(1)	& 4.1(1)	& mm \\
Axial width, $z_{\rm{rms}}$ &	2.1(1)	& 2.6(1)	& mm\\
MOT beam full diameter, $d_{\lambda}$ &	23 &	23 &	mm \\
MOT beam $1/e^2$ diameter &	14	& 14 &	mm \\
\hline
\end{tabular}
\caption{Comparison of SrF MOT parameters determined using the new and original \cite{Barry2014} trapping schemes. $^{*}$ Loss into $N=3$ may have contributed a non-negligible fraction of the total loss rate that determined the original MOT lifetime.}
\end{table}

\section{Further cooling of trapped molecules}
To decrease the temperature of magneto-optically trapped atomic samples, it is common to increase the magnitude of the laser detuning for a brief period (often while also zeroing the magnetic field gradient).  For alkali atoms, this gives additional cooling due to sub-Doppler forces, leading to dramatically lower temperatures \cite{Lett1988}. We have now observed that a similar brief period of increased negative detuning can reduce the temperature of our tightly-trapped MOT.

\begin{figure}
\centering
\includegraphics[width=10cm]{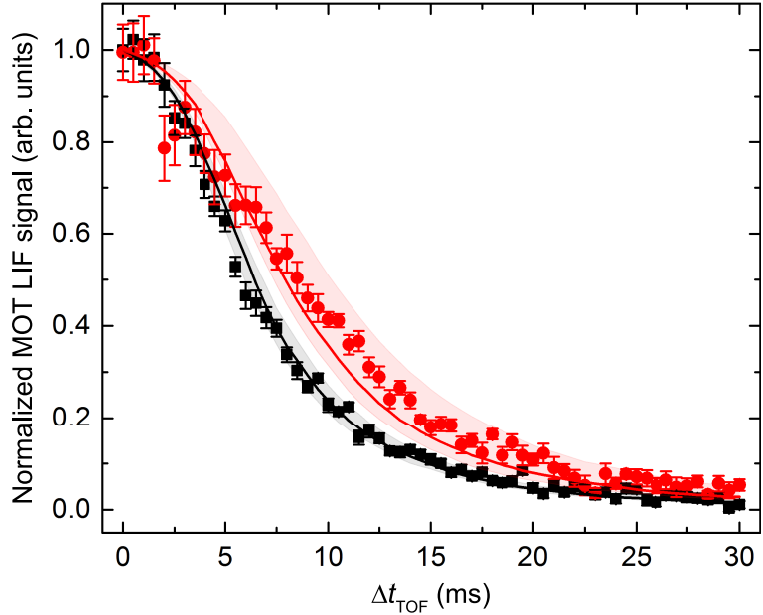}
        \caption{Further cooling of trapped molecules. Recaptured MOT fraction as a function of $\Delta t_{\rm{TOF}}$ for a fixed detuning of $\Delta=-10$~MHz (black squares) and a 10~ms duration of $\Delta= -20$~MHz immediately before release (red circles). Fits to the data are derived from Monte Carlo simulations and give upper limits on the isotropic temperature of 12(2)~mK and 8(2)~mK respectively. Shaded regions around the fits and error bars show the $\pm1\sigma$ confidence intervals based on a $\chi^{2}$ analysis.}
        \label{fig:Cooling}
\end{figure}

Using the release and recapture method, the temperature of the trapped molecules is measured when $\Delta$ is rapidly jumped from $-2\pi\times 10$~MHz $\approx-1.5\Gamma$ to $- 2\pi\times 20$~MHz $\approx-3\Gamma$ for 10~ms prior to release ($t=90$~ms to $t=100$~ms) before returning to $-2\pi\times 10$~MHz for recapture. The usual analysis indicates that this causes the temperature to decrease to $T_{\rm{iso}}<8(2)$~mK (Fig. \ref{fig:Cooling}). In addition, this method likely underestimates the temperature decrease due to the period of increased detuning. If the temperature were unchanged, one would expect a more rapid loss of molecules as a function of $\Delta t_{\rm{TOF}}$ due to the trapped cloud expanding during the 10~ms at increased detuning (and consequently reduced confining forces). The fact that molecules are measured to leave the trap volume at a slower rate despite this expansion prior to release strongly suggests that our analysis method overestimates the temperature. More than $80\%$ of the trapped molecules survive the 10~ms cooling stage.

Given our current understanding of sub-Doppler cooling mechanisms in this transition of SrF \cite{Shuman2010}, the decrease in temperature here cannot be due to the same mechanisms typically at play in alkali-atom MOTs. Instead, a plausible explanation for this cooling relies on the fact that our molecular MOT operates at unusually high laser intensity, and that heating and Doppler cooling rates vary quite differently as a function of laser detuning for a strongly saturated transition.

The need for very high laser intensity and hence strong saturation arises from the multi-level nature of the cycling transition \cite{Tarbutt2013}. In total, this transition includes 24 $X(v=0,1;N=1)$ ground-state sublevels (12 in each vibrational level $v=0,1$) and 4 $A(v=0,J=1/2)$ excited-state sublevels. At saturation, assuming complete remixing of ground-state sublevels, a molecule spends an equal amount of time in each sublevel and the maximum scattering rate is $R^{\rm{max}}_{\rm{sc}}\approx\Gamma \times \frac{4}{24+4}$  \cite{Tarbutt2013}. Additionally, the modest confining forces present in the MOT demand high laser intensity and a large scattering rate to produce a visible MOT.

To understand qualitatively how the heating and cooling rates can depend differently on detuning, consider a simple case of 1D excitation of a two-level system by counterpropagating laser beams. Here the scattering rate, $R^{\rm{2-level}}_{sc}$, and the damping coefficient, $\alpha^{\rm{2-level}}$, are given by \cite{Metcalf1999}
\begin{equation}
R^{\rm{2-level}}_{\rm{sc}}(\Delta)=\frac{\Gamma}{2}\frac{s}{\big(1+s+4(\Delta/\Gamma)^{2}\big)},\    \ \rm{and}
\label{eq:Scatt}
\end{equation}
\begin{equation}
\alpha^{\rm{2-level}}(\Delta)=- \hbar k^{2} \frac{4 s (\Delta/\Gamma)}{\big(1+s+4(\Delta/\Gamma)^{2}\big)^{2}},
\label{eq:Damp}
\end{equation}
where $s$ is the saturation parameter. For any value of $s$, the maximum total scattering rate (and hence maximum heating rate) occurs at zero detuning, $\Delta = 0$, so increasing the magnitude of the MOT laser detuning always decreases the scattering rate. In contrast, the maximum damping coefficient occurs at a finite (negative) detuning, $\Delta_d^{\mathrm{max}}$, and the change in the damping coefficient when moving to more negative detunings depends on the value of $s$.
The multi-level nature of our system makes it difficult to define $s$ exactly. However our measured value of the scattering rate, $R_{\rm{sc}}$, is near the maximum value $R^{\rm{max}}_{\rm{sc}}$; this in turn indicates that for any sensible definition, $s\gg1$ under our conditions. For sufficiently large values of $s$, the relative decrease in the scattering rate when moving to more negative detunings can be greater than the relative decrease in the damping coefficient. Since (in simple Doppler cooling) the heating rate is proportional to the scattering rate and the cooling rate is proportional to the damping coefficient, such a change would result in a decrease in temperature, as observed here. However, we have not attempted to make a quantitative estimate of the temperature change we expect when moving from $\Delta=-2\pi\times10$~MHz $\approx-1.5\Gamma$ to $\Delta=-2\pi\times20$~MHz $\approx-3.0\Gamma$ under our conditions.

\section{Conclusion}
The increased magneto-optical confinement of a diatomic molecule presented in this work is in good qualitative agreement with recent numerical simulations and physical arguments presented in ref. \cite{Tarbutt2014}. These results provide insight into the mechanisms behind the restoring force in type-II MOTs and suggest that the polarizations needed to generate confining forces in any type of MOT are, as proposed in ref. \cite{Tarbutt2014}, dictated only by the sign of the excited state g-factor, $g'$, and the ground and excited state total angular momenta, $F$ and $F'$. This understanding is imperative when magneto-optically trapping species with complex energy level structures such as those present in molecules.

The trapping scheme used in this MOT provides tighter confinement than could be achieved with our original configuration, with spring constants comparable to those reported in atomic type-II MOTs \cite{Tiwari2008} (though still one to two orders of magnitude less than those in typical type-I atomic MOTs \cite{Wallace1994}). We note that a MOT was observed in ref. \cite{Barry2014} when driving confining transitions from the $|J=3/2,F=2\rangle$ and $|J=1/2,F=1\rangle$ manifolds (but anti-confining transitions from the $|J=3/2,F=1\rangle$ and $|J=1/2,F=0\rangle$ manifolds), but not when the MOT polarizations or magnetic field gradient were reversed. This suggests that a significant fraction of the restoring force is due to the $F=2\rightarrow F'=1$ transition, in agreement with the arguments in ref. \cite{Tarbutt2014}. Although the measured increase in confinement between the original trapping scheme of ref. \cite{Barry2014} and the new scheme used here is in qualitative agreement with the analysis in ref. \cite{Tarbutt2014}, that work overestimates the cooling forces present in the MOT in ref. \cite{Barry2014} by more than an order of magnitude. This reveals that some work remains to be done to fully understand the behavior of type-II MOTs such as the one described in this paper. Nevertheless, the new understanding of type-II MOTs outlined in ref. \cite{Tarbutt2014} and supported by our results makes it appear that ordinary magneto-optical trapping methods could be applied to a significant number of diatomic species amenable to laser cooling \cite{Stuhl08,DiRosa04}, as long as $g'\neq0$ in the excited state of the cycling transition.

Future work is expected to give access to larger trapped molecular samples at higher density. For example, the flux of slow molecules passing through the trapping region may be increased by chirping the slowing laser frequency \cite{Zhelyazkova2014}, by slowing more efficiently via stimulated rather than spontaneous forces \cite{Chieda2011}, or by applying transverse confinement to the molecular beam during the slowing \cite{DeMille2013}. One method that may increase the trap depth employs a rapid synchronous reversing of the MOT beam circular polarizations and the magnetic field gradient (the RF MOT). This method of producing a restoring force was demonstrated in the two-dimensional magneto-optic compression of a molecular beam \cite{Hummon2013}, and according to the calculations of ref. \cite{Tarbutt2014} should provide 3D trapping forces comparable to those in atomic type-I MOTs, even when $g'=0$. However, the RF MOT technique is far more technically challenging to implement in 3D than the usual static MOT configuration. In addition, with properly chosen polarizations as prescribed in ref. \cite{Tarbutt2014}, the static type-II MOT should be just as effective as the RF MOT for transitions where $g'\sim1$, as for example in molecular species such as TlF \cite{Hunter2012}, AlF \cite{Wells2011}, and BH \cite{Hendricks2014} and atomic species such as In \cite{Kim2009a}. Hence we expect that this straightforward method will be useful for producing dense ultracold gases of diatomic molecules. This may enable access to diverse new experiments including searches for variations of fundamental constants \cite{Chin09}, studies of chemical dynamics at ultracold temperatures \cite{Meyer2011}, and tests of the standard model \cite{Hunter2012,Tarbutt2013}.

\section*{Acknowledgments}
We acknowledge funding from ARO and ARO (MURI). E.B.N acknowledges funding from the NSF GRFP.

\section*{References}
\bibliography{thebib}

\end{document}